\documentclass[twocolumn,aps,longbibliography]{revtex4-2}

\usepackage[pdftex]{graphicx}
\usepackage{amsmath}

\begin{document}

\title{Effects of doping on polar order in SrTiO$_{3}$ from first-principles modeling}

\author{Alex Hallett}
\affiliation{Materials Department, University of California, Santa Barbara, California 93106, USA}

\author{John W. Harter}
\email[Corresponding author: ]{harter@ucsb.edu}
\affiliation{Materials Department, University of California, Santa Barbara, California 93106, USA}

\date{\today}

\begin{abstract}
SrTiO$_{3}$ is an incipient ferroelectric and an exceptionally dilute superconductor with a dome-like dependence on carrier concentration. Stabilization of a polar phase through chemical substitution or strain significantly enhances the superconducting critical temperature, suggesting a possible connection between the polar instability and unconventional Cooper pairing. To investigate the effects of doping on the polar order in SrTiO$_{3}$, we develop a simplified free energy model which includes only the degrees of freedom necessary to capture the relevant physics of a doped, biaxially compressively strained system. We simulate the polar and antiferrodistortive thermal phase transitions using Monte Carlo methods for different doping levels and comment on the doping dependence of the transition temperatures and the formation of polar nanodomains. In addition, the temperature-dependent phonon spectral function is calculated using Langevin simulations to investigate the lattice dynamics of the doped system. We also examine the effects of doping on the electronic structure within the polar phase, including the density of states and band splitting. Finally, we compute the polarization dependence of the Rashba parameter and the doping dependence of the Midgal ratio, and place our results in the broader context of proposed pairing mechanisms. 
\end{abstract}

\maketitle

\section{Introduction}

SrTiO$_{3}$ becomes superconducting at exceptionally low carrier concentrations, resulting in a small Fermi energy ($E_F\sim 1$~meV) relative to the characteristic phonon frequency ($\omega_{D} \sim 100$~meV). The large Migdal ratio ($\omega_{D}/E_F$) is generally believed to place SrTiO$_{3}$ outside the adiabatic regime, rendering conventional Bardeen-Cooper-Schrieffer (BCS) theory inadequate to describe superconductivity in this system. Remarkably, stabilizing polar order through strain or isotope substitution has been shown to significantly enhance the superconducting critical temperature~\cite{edgeQuantumCriticalOrigin2015,stuckyIsotopeEffectSuperconducting2016,rischauFerroelectricQuantumPhase2017,ahadiEnhancingSuperconductivitySrTiO2019,russellFerroelectricEnhancementSuperconductivity2019,rischauFerroelectricQuantumPhase2017,rischauIsotopeTuningSuperconducting2022,hameedEnhancedSuperconductivityFerroelectric2022}, suggesting that unconventional Cooper pairing may be intimately related to the polar order. Despite decades of experimental and theoretical work, there is no consensus on the mechanism of superconductivity in SrTiO$_3$. There exist many posited theories, most of which involve coupling to critical fluctuations of the polar order parameter. Within this framework, superconductivity is enhanced as fluctuations intensify approaching the quantum critical point from the paraelectric phase, and diminishes as fluctuations subside within the ferroelectric phase~\cite{edgeQuantumCriticalOrigin2015,arce-gamboaQuantumFerroelectricInstabilities2018,koziiSuperconductivityFerroelectricQuantum2019,volkovSuperconductivityEnergyFluctuations2021, rowleySuperconductivityVicinityFerroelectric,enderleinSuperconductivityMediatedPolar2020,kedemNovelPairingMechanism2018, rowleyFerroelectricQuantumCriticality2014,rischauFerroelectricQuantumPhase2017,rischauIsotopeTuningSuperconducting2022,hameedEnhancedSuperconductivityFerroelectric2022,fauqueMesoscopicTunnelingStrontium2022}. Specific proposed mediators of pairing include a single transverse optical (TO) phonon mode~\cite{yuTheorySuperconductivityDoped2021,yoonLowdensitySuperconductivitySrTiO2021,gastiasoroTheorySuperconductivityMediated2022,zyuzinAnisotropicResistivitySuperconducting2022}, exchange between two TO phonons~\cite{ngaiTwoPhononDeformationPotential1974,vandermarelPossibleMechanismSuperconductivity2019,kiseliovTheorySuperconductivityDue2021,zyuzinAnisotropicResistivitySuperconducting2022,gastiasoroGeneralizedRashbaElectronphonon2023}, and exchange between longitudinal optical (LO) modes~\cite{gorkovPhononMechanismMost2016,gastiasoroPhononmediatedSuperconductivityLow2019}. 

While there are many studies that support the quantum critical theories, recent work has uncovered the existence of polar nanodomains at high temperatures, even in unstrained films~\cite{salmani-rezaieOrderDisorderFerroelectricTransition2020}, as well as enhanced superconductivity deep within the polar phase~\cite{russellFerroelectricEnhancementSuperconductivity2019}, where critical fluctuations of the order parameter are negligible. These experiments have motivated the search for alternative theoretical paradigms to explain the pairing mechanism in SrTiO$_3$, such as the inversion symmetry breaking framework. In this class of theories, Cooper pairs effectively exist in a noncentrosymmetric environment, even in films that do not exhibit global polar order, such as unstrained or highly doped films. Superconductivity persists so long as the length scale of the nanodomains is longer than the coherence length of the Cooper pairs. Furthermore, a direct correlation has been found between nanodomain size and the superconducting and polar transition temperatures in doped, compressively strained SrTiO$_3$. As nanodomains are destroyed by nonmagnetic~\cite{salmani-rezaiePolarNanodomainsFerroelectric2020,salmani-rezaieRoleLocallyPolar2021} or magnetic dopants~\cite{salmani-rezaieSuperconductivityMagneticallyDoped2021}, films that do not have sufficiently large nanodomains at room temperature do not undergo a superconducting transition. These studies present strong evidence for inversion symmetry breaking as a key prerequisite for superconductivity. The specific pairing mechanism in this scenario remains unknown, but the existence of unconventional superconductivity is supported by nonreciprocal charge transport~\cite{schumannPossibleSignaturesMixedparity2020} and insensitivity to magnetic dopants~\cite{salmani-rezaieSuperconductivityMagneticallyDoped2021}. 

Another proposed mechanism involves Rashba coupling, which combines aspects of both the quantum critical and inversion symmetry breaking frameworks~\cite{yuTheorySuperconductivityDoped2021,gastiasoroGeneralizedRashbaElectronphonon2023,gastiasoroTheorySuperconductivityMediated2022,yoonLowdensitySuperconductivitySrTiO2021}. The suggested mediator of pairing is a single, soft TO polar phonon mode in the presence of strong spin-orbit coupling and inversion symmetry breaking. Coupling between the electrons and TO phonons is typically prohibited because these modes do not modulate the charge density. Nevertheless, the polar TO phonon mode breaks inversion symmetry, and in the presence of spin-orbit coupling leads to a Rashba spin splitting of the conduction bands, allowing for an additional hopping channel between neighboring $d$ orbitals. Prior work, however, indicates that the Rashba coupling in SrTiO$_3$ may not be sufficiently strong to account for superconductivity away from the quantum critical point~\cite{gastiasoroTheorySuperconductivityMediated2022}, where critical fluctuations increase the electron-phonon coupling strength. Thus, this theory aligns more with the quantum critical framework, and therefore does not explain the experimental evidence of enhanced superconductivity deep within the polar phase in compressively strained system.

In this work, we expand our previously-described Landau free energy model~\cite{hallettModelingPolarOrder2022} to include the effects of electron doping in compressively strained SrTiO$_3$, and use statistical mechanics methods to simulate thermal disorder in large supercells. Monte Carlo simulations are used to calculate transition temperatures and domain structures for different doping levels. Langevin dynamics is implemented to simulate the temperature dependence of the phonon spectral function in systems with varying carrier concentrations, and to compute the phonon frequencies of the polar modes within the ordered phase. The electronic structure of the tetragonal I4/mcm phase is compared to that of the strained I4cm phase, where octahedral rotations coexist with polar distortions. Finally, we calculate the effects of doping on the Rashba parameter and the Migdal ratio, and discuss our results as they relate to existing theories of superconductivity in SrTiO$_3$.

\begin{figure*}[t]
\includegraphics{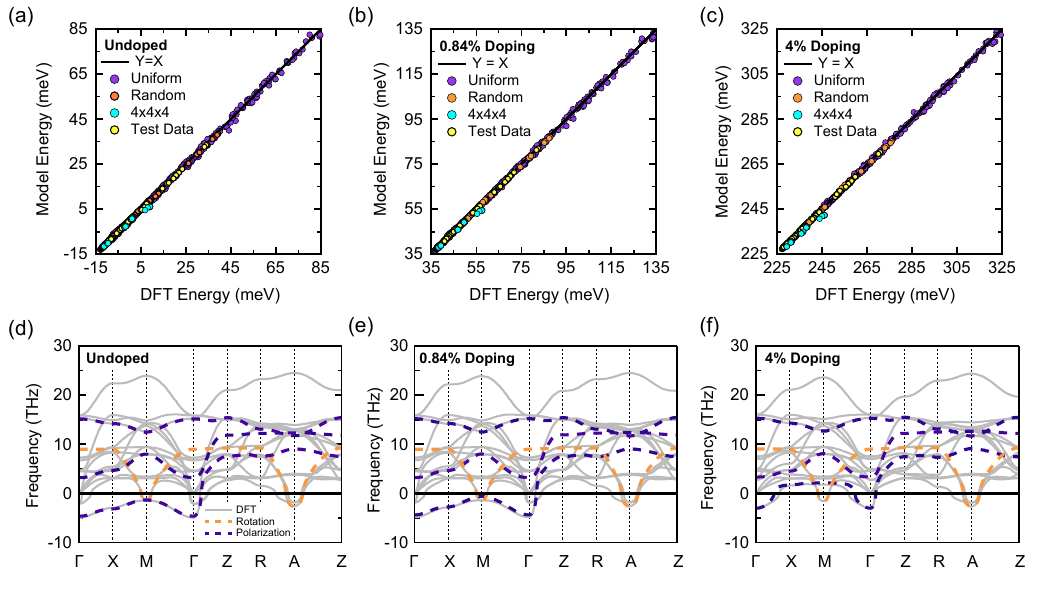}
\caption{\textbf{Validation of the free energy model at zero temperature.} Model versus DFT energies of 1392 distinct configurations for (a)~no doping, (b)~0.84\% doping, and (c)~4\% doping. The model energies are in excellent agreement with DFT, with root mean square errors of 0.16, 0.14, 0.12~meV/atom, respectively. (d)-(f)~The four low-energy phonon bands (1 rotation, 3 polarization) for the strained centrosymmetric reference structure calculated for the model and overlaid on the phonon dispersion calculated using DFT, showing that the simple free energy model can capture the relevant instabilities with near DFT-level accuracy for all three doping levels.}
\label{fig:Fig_1_Free_Energy}
\end{figure*}

\section{Computational Procedure}

\subsection{Ground State Structures}

The ground state structure of undoped compressively strained SrTiO$_3$ was calculated through a series of structural relaxations described in our prior work~\cite{hallettModelingPolarOrder2022}. Calculations were performed using density functional theory (DFT) as implemented in the Vienna ab intio simulation package (\textsc{vasp})~\cite{kresseInitioMolecularDynamics1993, kresseEfficientIterativeSchemes1996, kresseEfficiencyAbinitioTotal1996}. The DFT parameters for this work are identical to those in Ref.~\citenum{hallettModelingPolarOrder2022} and can also be found in the Supplemental Materials~\cite{SeeSupplementalMaterial}.

Compressive in-plane strain values were chosen to replicate SrTiO$_3$ thin films grown on LSAT. To improve the accuracy of transition temperatures and compensate for the overestimation of the $c$-axis lattice parameter by the GGA approximation~\cite{tranRungsDFTJacob2016}, the out-of-plane strain was fixed at the experimental value of 0.71\%~\cite{yamadaPhaseTransitionsAssociated2015}. More details on the out-of-plane strain can be found in the Supplemental Materials~\cite{SeeSupplementalMaterial}. To examine the effects of doping, the carrier concentration was modulated by increasing the total number of valence electrons within the unit cell. A compensatory positive background charge was applied to maintain overall charge neutrality. Calculations were performed for three characteristic doping levels: undoped (0\%), $1.4 \times 10^{19}$~cm$^{-3}$ (0.84\%), and $6.63 \times 10^{20}$~cm$^{-3}$ (4\%). The numerical values of the ground state atomic displacements for each doping level are given in Table~\ref{tab:Tab_1}.  

In subsequent discussions, order parameter amplitudes are defined by the ion displacements in the undoped ground state structure relative to the strained, centrosymmetric reference state. According to this definition, all order parameter amplitudes vanish in the reference state. In the ground state, order parameters are normalized by the displacements in the undoped system so that the magnitudes of displacements for each doping level can be compared. The global polarization order parameter comprises three individual degrees of freedom of the polar distortion: the titanium displacements, the in-plane oxygen displacements, and the out-of-plane oxygen displacements. The net polarization is calculated as the component of the titanium and oxygen ion displacement vector along the direction of the ground state displacement vector. The rotation order parameter is the absolute value of the in-plane displacement of the oxygen atoms, accounting for averaging between neighboring unit cells with different rotation amplitudes. For the ground state structures, the normalized polarization order parameters from lowest to highest doping are 1.000, 0.627, and 0.331, while the normalized ground state rotation order parameters are 1.000, 1.004, and 1.095, which correspond to 4.99$^\circ$, 5.01$^\circ$, and 5.38$^\circ$ octahedral rotation angles, respectively. The rotation angle increases as the polar order is suppressed with doping, which has been explored in more detail in our previous work~\cite{hallettModelingPolarOrder2022}. The explicit definition of the order parameters can be found in the Supplemental Materials~\cite{SeeSupplementalMaterial}. 
 
\begin{table}
\caption{Ground state distortions (\AA).}
\begin{tabular}{c c c c}
\hline
\hline
Ion Type \& Direction & 0\% & 0.84\% & 4\%	\\
\hline
Titanium ($\hat{z}$)				&	$0.0305$	& $0.0299$	 & 	$0.0206$ \\
In-Plane Oxygen ($\hat{z}$)			&	$-0.0697$	& $-0.0682$	 & 	$-0.0390$ \\
Out-of-Plane Oxygen ($\hat{z}$)		&	$-0.0749$	& $-0.0731$	 & 	$-0.0438$ \\
Rotation ($\hat{x}$/$\hat{y}$)	    &	$0.1702 $	& $0.1708$	 &	$0.1863$  \\ 
\hline
\hline
\end{tabular}
\label{tab:Tab_1}
\end{table}

\subsection{Free Energy Model} 

DFT is limited by its inability to accommodate thermal effects and by the considerable computational expense associated with large, disordered systems. To simulate thermal phase transitions in SrTiO$_3$, we developed a simple model that can efficiently incorporate both finite temperature and associated disorder. Following our prior work~\cite{hallettModelingPolarOrder2022}, we approximate the free energy of the system in accordance with Landau theory, performing a Taylor series expansion about the relevant order parameters and yielding a sum of invariant polynomials. In formulating the free energy expression, we consider only four degrees of freedom: the three components of the polarization (titanium and in- and out-of-plane oxygen ions), and the octahedral rotations. The \textsc{isotropy} software suite~\cite{stokesISOTROPYSoftwareSuite,Hatch:wt0012} was used to find all symmetry-allowed terms up to fourth order in rotation and polarization, including couplings of the order parameters to their 26 nearest-neighbors. Strictly on-site terms were included up to sixth order to prevent divergences. The final free energy polynomial consists of 69 distinct terms, with the full expression given in the Supplemental Materials~\cite{SeeSupplementalMaterial}.  

After identifying the allowed terms in the free energy, their coefficients were determined using high-throughput DFT calculations of $2 \times 2 \times 2$ supercell configurations, choosing either random or uniform values for each degree of freedom in each unit cell. The different types of configurations are listed in the Supplemental Materials~\cite{SeeSupplementalMaterial}. The same total set of configurations was used for all three doping levels, yielding three separate sets of model coefficients. A total of 1392 configurations were considered in calculating the model parameters for each doping level. After calculating the free energy coefficients using least squares linear regression, we used the model to calculate the energies of 84 additional test configurations, to confirm the generality of our free energy expression. The error in the calculated energies for these test structures was comparable to that of the training data set. We also calculated the energies of larger $4 \times 4 \times 4$ supercells to determine if excluding longer-range interactions affected the accuracy of the model. The model energies are plotted versus the DFT energies for each doping level in Fig.~\ref{fig:Fig_1_Free_Energy}(a)-(c). The root mean square errors of the model energies relative to DFT were 0.16, 0.14, and 0.12 meV/atom, from lowest to highest doping. While the $4 \times 4 \times 4$ supercell energies deviate from the model for high-energy configurations, they are accurately calculated by the model close to the ground state, which is most relevant for simulations at reasonable temperatures. The discrepancy between the DFT and model energies for the larger supercells far above the ground state could be due to domain walls in the rotation order parameter or to increased coupling between next-nearest-neighbors at high displacement amplitudes. 

As a final check, we compared the phonon dispersion of our free energy model to that calculated by DFT. A derivation of the phonon dispersion for our model is provided in the Supplemental Materials~\cite{SeeSupplementalMaterial}. Figures~\ref{fig:Fig_1_Free_Energy}(d)-(f) show the model and DFT phonon dispersions for each doping level. We do not expect the model to accurately capture the high-frequency bands, as we do not consider all phonon degrees of freedom. Our goal is merely to capture the relevant structural instabilities accurately, with a dramatically reduced phase space volume. As shown in Fig.~\ref{fig:Fig_1_Free_Energy}, we successfully replicate the low-energy phonon bands with near DFT-level accuracy.

\section{Results and Discussion}

\subsection{Polar Nanodomain Stability} 

\begin{figure}[b]
\includegraphics{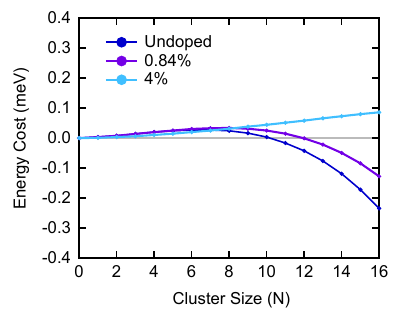}
\caption{\textbf{Stability of polar clusters.} Energy cost versus cluster size for $N \times N \times N$ polar domains embedded in an unpolarized background. The uniform rotation order parameter is set to the ground state value for each doping level.}
\label{fig:Fig_2_Cluster}
\end{figure}

After validating the model's accuracy, we assessed the stability of polar nanodomains across different doping levels. Our results reveal that while abrupt domain walls between oppositely polarized domains incur a significant energy cost, it can in fact be energetically favorable for polar nanodomains to form within an unpolarized background. This unpolarized reference state is likely to exist, at least on average, at temperatures far above the polar transition. We performed zero-temperature calculations for polar clusters of varying dimensions to explore the energetics of nanodomain formation within a reference state where the polar order parameter is zero and the octahedral rotation is set to the ground state value. This represents a system after the antiferrodistortive (AFD) transition but before the polar transition. Inside the cluster, the magnitude of the polarization was set to the ground state value corresponding to each doping level.

Figure~\ref{fig:Fig_2_Cluster} shows energy versus cluster size calculated using our free energy model. When the system energy becomes negative, the formation of domains is favorable. For an $N \times N \times N$ domain, this occurs when $N = 10$ for the undoped system and $N = 12$ for the 0.84\% system. For the highest doping level, we did not observe a stable nanodomain within the range of values of $N$ that we tested. In all cases, the energy increases initially because of the cost of the domain wall ($\propto N^2$), but becomes negative when the cluster reaches a critical size due to the energy lowering within the bulk of the nanodomain ($\propto -N^3$). These results are in general agreement with experimental observation showing that increased doping destroys polar nanodomains in SrTiO$_3$~\cite{zhuCoexistenceAntiferrodistortivePolar2024}. 

\subsection{Simulating Thermal Phase Transitions}

\begin{figure*}[htbp]
\includegraphics{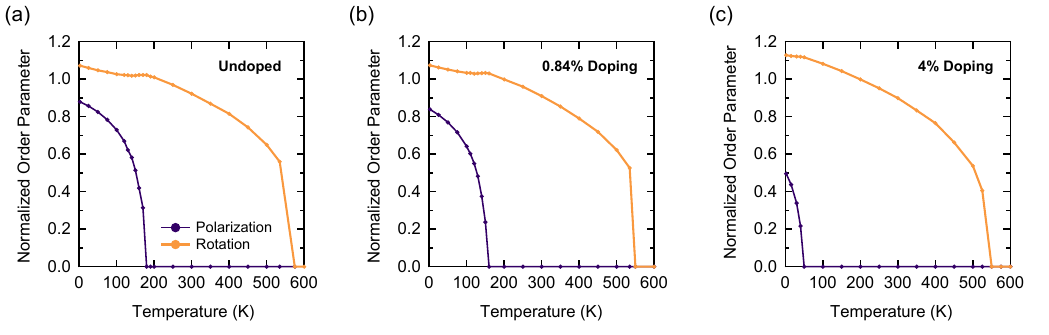}
\caption{\textbf{Simulation of thermal phase transitions.} Rotation and polarization order parameters versus temperature. (a)~For the undoped system, the polar transition occurs at 180~K and the AFD transition occurs at 560~K. (b)~For the 0.84\% doped system, the polar transition occurs at 160~K and the AFD transition occurs at 550~K. (c)~For the 4\% doped system, the polar transition occurs at 50~K and the AFD transition occurs at 550~K. The slight kink in the rotation order parameter at the polarization transition is due to coupling between the order parameters. As the polarization rapidly decreases, the rotation angle increases. The competition between these order parameters is discussed extensively in Ref.~\citenum{hallettModelingPolarOrder2022}.}
\label{fig:Fig_3_Monte_Carlo}
\end{figure*}

We used the Metropolis Monte Carlo algorithm in combination with our free energy model to simulate the temperature-dependent polar and AFD phase transitions. We considered thermal fluctuations of all four degrees of freedom in our model (three components of the polarization together with octahedral rotations) within a large 16 $\times$ 16 $\times$ 16 supercell with periodic boundary conditions. The thermally-averaged order parameters are plotted versus temperature in Fig.~\ref{fig:Fig_3_Monte_Carlo}. Technical details of the Monte Carlo simulations can be found in the Supplemental Materials~\cite{SeeSupplementalMaterial}.

Table~\ref{tab:Tab_2} compares our results to experimental studies of strained, doped films. This table includes the doping levels, in-plane strain ($\epsilon_{\parallel}$), out-of-plane strain ($\epsilon_{\perp}$), room-temperature experimental lattice parameters, and polar ($T_\mathrm{P}$) and AFD ($T_\mathrm{AFD}$) transition temperatures. Our procedure for finding the in- and out-of-plane strain values can be found in the Supplemental Materials~\cite{SeeSupplementalMaterial}. The measurements or calculations of strain from other works listed in Table~\ref{tab:Tab_2} can be found in the corresponding references. The simulated polar transition temperature of the undoped system is within 25~K of the experimentally measured value, which is in remarkable quantitative agreement considering the simplified model that we use. In addition, the transition temperature decreases significantly with doping, which is in qualitative agreement with experiment~\cite{russellFerroelectricEnhancementSuperconductivity2019}. 

The overestimation of the $c$-axis lattice constant, and in particular the additional elongation upon transitioning from the centrosymmetric to the polar phase, was a possible source of error in previous work~\cite{hallettModelingPolarOrder2022}. Indeed, we found that when the out-of-plane strain was fixed at the experimental value, the polar transition temperature was reduced by 100~K for the undoped system, bringing it much closer to the experimental value. An alternative solution to the $c$-axis problem could be provided by a different functional, such as the the strongly constrained and appropriately normed (SCAN) functional, which has been shown to give accurate energies and structural parameters for perovskite oxides. The remaining discrepancies in the transition temperatures may be caused by the omission of anharmonic coupling effects between the low energy bands and higher energy phonons, which are neglected in our work. We also recognize that previous studies have found long-range dipole--dipole interactions to be significant, but which are much more computationally expensive than our simplified model~\cite{zhongFirstprinciplesTheoryFerroelectric1995}. 

\begin{table}
\caption{Comparison of experimental and computational transition temperatures for different doping levels. Entries labeled with an asterisk correspond to this work.}
\begin{tabular}{c c c c c c c c c}
\hline
\hline
Doping  & $\epsilon_{\parallel}$ (\%) & $\epsilon_{\perp}$ (\%) & $a$,$b$ (\AA) & $c$ (\AA) & $T_\textrm{P}$ (K) & $T_\textrm{AFD}$ (K) & Ref. \\
\hline
0\% 	  & $-0.9$                      & 0.8                     & --            & --   	    & 140		              & 360		               & \cite{yamadaAntiferrodistortiveStructuralPhase2010} \\
0\% 	  & $-0.92$                     & 0.71                    & 3.869         & 3.933	    & 155		              & 370		               & \cite{yamadaPhaseTransitionsAssociated2015} \\
0.36\% 	& $-1$                        & --                      & --            & --	      & 92		              & --	                 & \cite{russellFerroelectricEnhancementSuperconductivity2019} \\
0.84\% 	& $-1$                        & --                      & --            & --	      & 39		              & --	                 & \cite{russellFerroelectricEnhancementSuperconductivity2019} \\
1.6\%	  & $-1$                        & --                      & --            & --   	    & None		            & --		               & \cite{russellFerroelectricEnhancementSuperconductivity2019} \\
\hline
0.00\%  & $-1.00$                     & 0.71                    & 3.8995        & 3.967 	  & 180		              & 560		               & * \\
0.84\%	& $-1.00$	                    & 0.71                    & 3.8995        & 3.967	    & 160		              & 550		               & * \\
4.00\% 	& $-1.00$                     & 0.71                    & 3.8995        & 3.967 	  & 60		              & 550		               & * \\  
\hline
\hline
\end{tabular}
\label{tab:Tab_2}
\end{table}

To examine the softening of the polar mode as a function of temperature and doping, we directly calculated the temperature-dependent phonon spectral function using Langevin dynamical simulations. The Langevin equations of motion in momentum space are given by
$${\Ddot{\phi}_k = -M_k^{-1}K_k\phi_k - \frac{\partial V_\mathrm{NL}}{\partial x_k} -\gamma \dot \phi_k + M_k^{-\frac{1}{2}}\sqrt{2\gamma k_B T}\eta,}$$
where $\phi_k$ is a 4-vector containing the values of the order parameters at each $k$ point, $M_k$ is a ``mass'' matrix, $K_k$ is a ``spring constant'' matrix capturing linear restoring forces, $ V_\mathrm{NL}$ captures all remaining (nonlinear) forces, $\gamma$ is a phenomenological damping parameter, and $\eta$ is Gaussian-distributed thermal noise. More details can be found in the Supplemental Materials~\cite{SeeSupplementalMaterial}. 

\begin{figure*}[htbp]
\includegraphics[width=\textwidth]{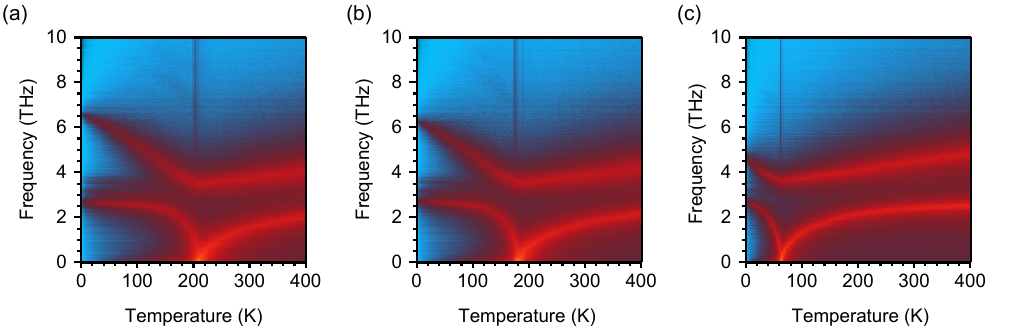}
\centering
\caption{\textbf{Polar phonon spectral functions.} Temperature dependence of the polar phonon spectral function at the $\Gamma$ point for (a)~the undoped system, (b)~0.84\% doping, and (c)~4.0\% doping. The vertical lines near each transition temperature are due to simulation noise from critical fluctuations.}
\label{fig:Fig_4_Langevin}
\end{figure*}

Starting with the ground state structure for a 16 $\times$ 16 $\times$ 16 supercell, we integrated the equations of motion as the system was ramped from high (500 K) to low (1 K) temperature. The spectral functions of the polar modes at the $\Gamma$ point are plotted as a function of temperature in Fig.~\ref{fig:Fig_4_Langevin} for each doping level. The lowest energy polar mode is observed to soften to zero frequency at the transition temperature, which is indicative of a displacive transition. The polar transition temperatures are 200~K, 180~K, and 60~K, in close quantitative agreement with our Monte Carlo simulations.

While one optical study has shown complete softening of the polar mode~\cite{takesadaPerfectSofteningFerroelectric2006}, many others show only incomplete softening~\cite{cowleyTemperatureDependenceTransverse1962,cowleyLatticeDynamicsPhase1964, barkerTemperatureDependenceTransverse1966, yamadaNeutronScatteringNature1969,shiraneLatticeDynamicalStudy1101969, buerleSoftModesSemiconducting1980, courtensPhononAnomaliesSrTiO1993,sirenkoSoftmodeHardeningSrTiO32000, nuzhnyyInfraredPhononSpectroscopy2011,nuzhnyyInfraredPhononSpectroscopy2011, inoueStudyStructuralPhase1983,vogtRefinedTreatmentModel1995, yamanakaEvidenceCompetingOrderings2000,ostapchukOriginSoftmodeStiffening2002, akimovElectricFieldInducedSoftModeHardening2000, shigenariRamanSpectraFerroelectric2003, rischauFerroelectricQuantumPhase2017,rischauIsotopeTuningSuperconducting2022,uweStressinducedFerroelectricitySoft1976,rowleySuperconductivityVicinityFerroelectric,enderleinSuperconductivityMediatedPolar2020}. The complete softening ($\omega \rightarrow 0$) observed in our simulations may be a result of the uniformity of the added charge density, as we are not accounting for any local strain effects or disorder introduced by the dopant atoms. Additionally, reducing the film thickness is shown to lead to hardening of the polar mode~\cite{katayama2008}, but we are simulating an infinite bulk system rather than a thin film. Nevertheless, we observe favorable nanodomain formation, which is clear evidence of a mixed character transition (displacive and order-disorder). This classification of the polar transition is relevant because it informs whether Cooper pairing is likely mediated by quantum critical fluctuations of the polar mode. Binary classifications can be limited in their descriptive power, and it is important to look at the nuances of the specific electronic and structural degrees of freedom of the system. Within the polar phase, the soft mode frequency hardens again to $\omega_1 \approx 2.5 $~THz for the lowest frequency mode and $\omega_2 \approx 6.5$~THz for the second polar mode. A third polar mode at $\omega_3 \approx 10$~THz is also calculated, but is not shown in Fig.~\ref{fig:Fig_4_Langevin} for clarity. Experimentally, enhanced superconductivity is observed deep within the polar phase in compressively strained SrTiO$_3$, where this mode clearly has a finite frequency. This calls into question the relevance of quantum critical fluctuations of the soft mode as the mediator of Cooper pairing. 

\begin{figure*}[t]
\includegraphics[width=1\textwidth]{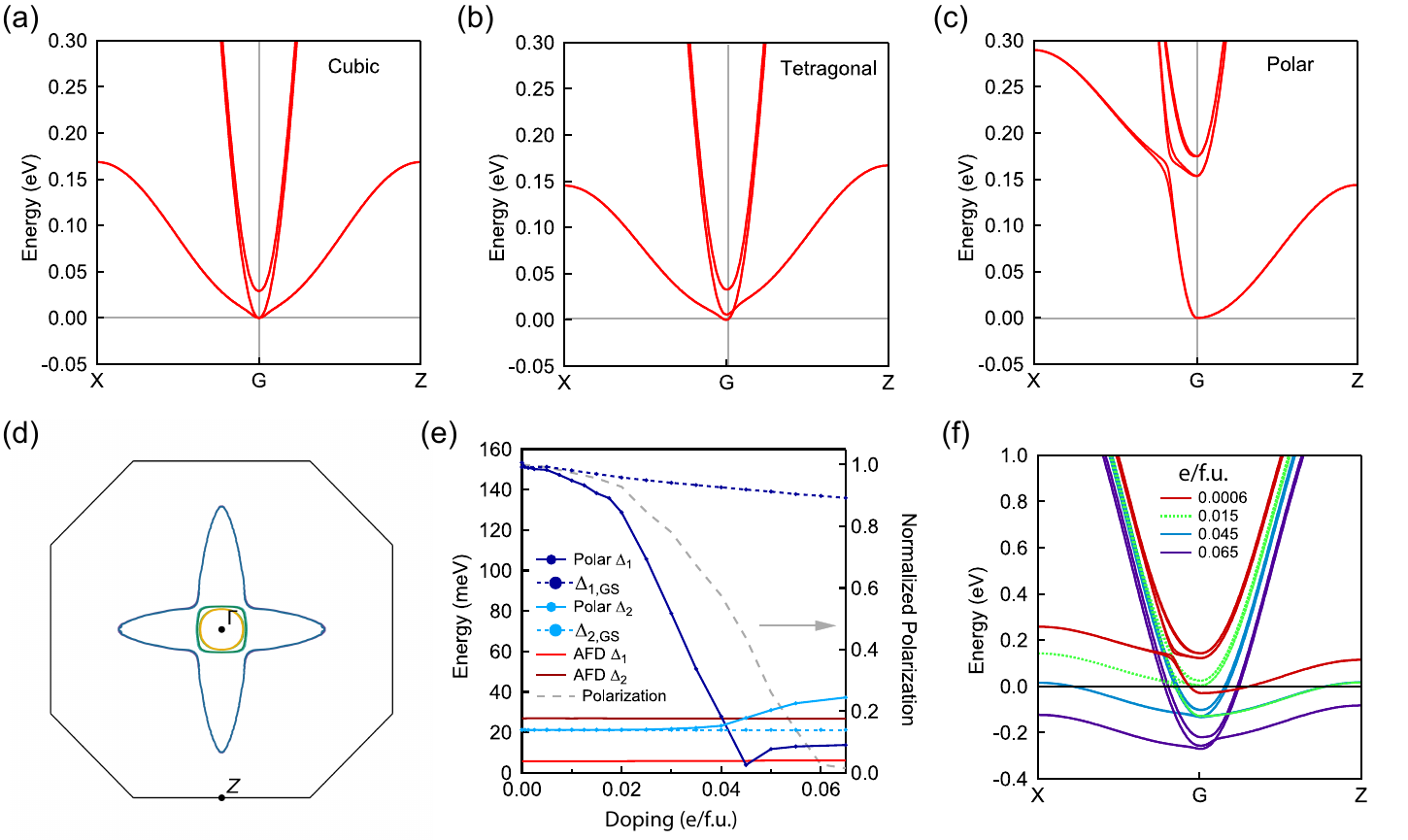}
\centering
\caption{\textbf{Electronic band structure.} Band structures for the undoped (a)~cubic (Pm$\bar{3}$m), (b)~tetragonal (I4/mcm), and (c)~polar (I4cm) structures along the path $X \rightarrow \Gamma \rightarrow Z$. (d)~Two dimensional slice of the Fermi surface for 4\% doping. The blue, green, and gold Fermi surface sheets correspond to lower, middle, and upper $t_{2g}$ bands, respectively. The topology of the Fermi surface for the I4cm polar structure is similar to the I4/mcm tetragonal phase. (e)~Energy splitting of the bands versus doping. While the energy splitting remains constant for the unstrained tetragonal structure (AFD $\Delta_{1,2}$), the splitting due to the polar distortion (Polar $\Delta_1$) changes drastically with doping. Blue dashed traces represent the energy splitting for the structures with polarization fixed at the ground state value. The polarization of the relaxed ground state structures for various doping levels (gray dashed line) is plotted against the right axis. (f)~Band structures for different doping levels, showing the decrease in band splitting at higher doping levels as the polar distortion is suppressed.}
\label{fig:Fig_5_band}
\end{figure*}

\subsection{Electronic Structure}

We now investigate the effects of doping on the electronic structure of SrTiO$_3$, with spin-orbit coupling included in all calculations. Band structures for the Pm$\bar{3}$m (cubic), I4/mcm (tetragonal), and I4cm (polar) phases are shown in Fig.~\ref{fig:Fig_5_band}(a)-(c). Undoped SrTiO$_3$ is an insulator with an indirect gap between filled oxygen $2p$ states and three unoccupied titanium $t_{2g}$ orbitals~\cite{gastiasoroSuperconductivityDiluteSrTiO2020}. Spin-orbit coupling causes a splitting between a lower $j=3/2$ quadruplet and a higher $j=1/2$ doublet in the high-temperature Pm$\bar{3}$m cubic structure, as shown in Fig.~\ref{fig:Fig_5_band}(a). At lower temperatures, the AFD octahedral rotations cause a transition to a tetragonal I4/mcm structure, splitting the degeneracy of the $j=3/2$ quadruplet and yielding three doublets at the $\Gamma$ point, as shown in Fig.~\ref{fig:Fig_5_band}(b). The lowest-energy doublet has a distinct non-parabolic dispersion, beginning as a light band but becoming heavier for $k > 0.1/a$ due to an avoided crossing~\cite{collignonMetallicitySuperconductivityDoped2019}. In the undoped, compressively strained system, shown in Fig.~\ref{fig:Fig_5_band}(c), the ground state polar distortion causes the lowest band to shift downward in energy substantially. A two-dimensional slice of the resulting Fermi surface for $n = 0.04$~e/f.u. is shown in Fig.~\ref{fig:Fig_5_band}(d). The Fermi surface of the polar structure evolves with doping similarly to that of bulk I4/mcm SrTiO$_3$~\cite{vandermarelCommonFermiliquidOrigin2011}: two bands as anisotropic spheres, and one band as three elongated ellipsoids along $x$, $y$, and $z$. The primary difference between the polar and low-temperature tetragonal electronic structures is the location of the Lifshitz transition due to the large shift of the lowest-energy titanium $t_{2g}$ band, together with an elongation along $z$ of the lobes of the outermost Fermi surface due to the additional strain. 

\begin{figure*}[htbp]
\centering
\includegraphics[width=1\textwidth]{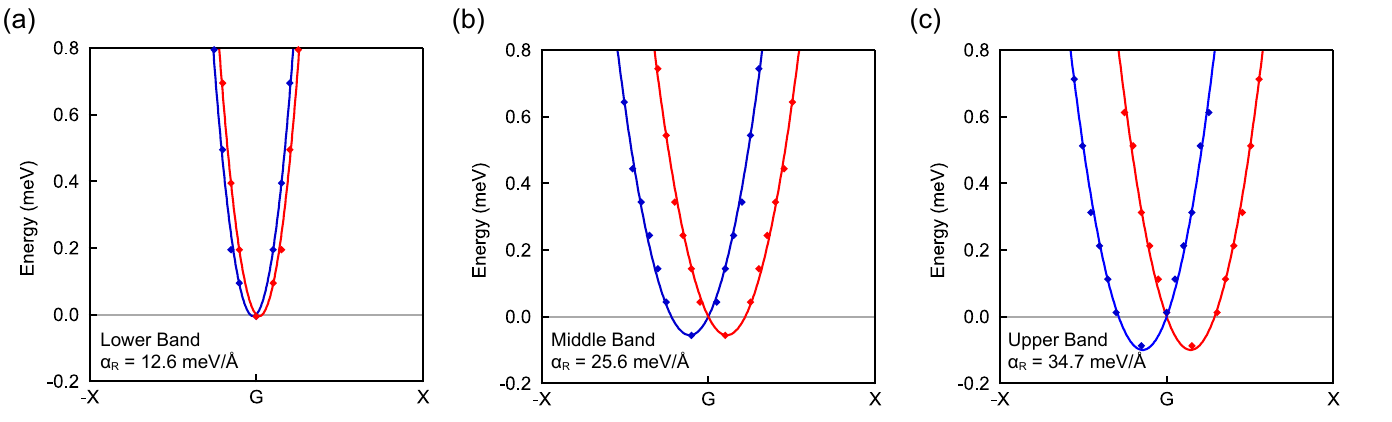}
\caption{\textbf{Rashba splitting.} Calculated Rashba band splittings for the (a)~lower, (b)~middle, and (c)~upper $t_{2g}$ bands of the undoped polar structure. The splitting increases in magnitude from the lowest to the highest energy band.}
\centering
\label{fig:Fig_6_rashba_splitting}
\end{figure*} 

The energy shift of the lowest conduction band is strongly dependent on the magnitude of the polar distortion, which in turn depends on the level of doping, as illustrated in Fig.~\ref{fig:Fig_5_band}(e). Energy splittings $\Delta_1$ and $\Delta_2$ are defined as the difference in energy between the lowest and middle bands, and the middle and upper bands, respectively. These quantities are shown for both the compressively strained polar structure (blue traces) and the unstrained structure with rotations (red traces). The doping dependence of the energy splitting when the polarization is fixed at the ground state value (blue dashed traces) is also shown to distinguish between the effects of increased carrier concentration and those of the structural changes caused by doping. 

For the unstrained AFD structure, the bands rigidly shift with increasing carrier concentration, and the band splittings therefore remain nearly constant at $\Delta_1 \approx 6$~meV and $\Delta_2 \approx 27$~meV. For the strained, polar structure, $\Delta_1 \approx 150$~meV when undoped, but begins to rapidly decrease near $n = 0.02$~e/f.u. as the polar distortion is suppressed. The second energy splitting $\Delta_2$, on the other hand, remains constant until $n = 0.045$~e/f.u., where it increases slightly. At the highest doping levels, where the polar distortion is completely suppressed, the energy splittings for the polar structure are slightly higher than those for the AFD structure, likely due to the increased elongation of the $c$-axis in the compressively strained system. For structures with a fixed polarization, the first energy gap $\Delta_{1,\mathrm{GS}}$ undergoes a linear decrease of $\sim$~15~meV across the entire doping range, while $\Delta_{2,\mathrm{GS}}$ remains constant. This demonstrates that while the increased carrier density slightly lowers the energy gap between the lowest and middle conduction bands, the primary effect on $\Delta_{1}$ originates from the structural suppression of polarization with increased doping. The polarization of the relaxed structures is plotted along the right axis in Fig.~\ref{fig:Fig_5_band}(e) to show the relationship between $\Delta_{1}$ and the suppression of the polar distortion with increasing carrier concentration.

The band structures at several doping levels are shown in Fig.~\ref{fig:Fig_5_band}(f), depicting the change with doping as the magnitude of the polar distortion decreases and the band structure approaches that of the strained, centrosymmetric structure. The Fermi energy is extracted from density of states calculations and subtracted from the energies of the bands, such that $E_F = 0$ for each doping level. As doping increases, the gap between the lower and middle bands decreases while the energy splitting between the middle and upper bands ($\Delta_2$) remains relatively constant. At $n = 0.045$~e/f.u., the middle band approaches the lowest energy band, and both $\Delta_{1}$ and $\Delta_2$ increase again as the polarization is completely suppressed. 

\subsection{Rashba Splitting} 

The coexistence of inversion symmetry breaking and strong spin-orbit coupling can give rise to a bulk Rashba spin-orbit interaction, where the spin of a charge carrier is locked to its momentum.  An effective momentum-dependent electric field arises from the variation in the potential felt by the electron due to the symmetry of the underlying lattice. As a consequence of this inversion-asymmetry-induced electric field gradient, electrons experience a momentum-dependent magnetic field, lifting the band degeneracy. Specific spin textures differ depending on lattice symmetry. Rashba spin-orbit coupling originates from planar inversion asymmetry. For inversion symmetry breaking along the $z$ direction, the effective Hamiltonian has the form
$${H_\mathrm{RSOC} = \alpha_R(k_x\sigma_y - k_y\sigma_x),}$$
leading to two Fermi surface contours with opposite helical spin texture. The energy of the bands is given by
$${E_{\pm}(\vec{k}) = \frac{h^2k^2}{2m^{*}} \pm \alpha_R|\vec{k}|.}$$
The Rashba parameter is defined as $\alpha_R = \frac{2\Delta E_R}{k_R}$~\cite{varignonUnexpectedCompetitionFerroelectricity2024}, where $\Delta E_R$ is the difference between the band minima and the energy at which the bands cross and $k_R$ is the wavevector of the band minima relative to the $\Gamma$ point. We explore the magnitude and polarization dependence of the Rashba parameter $\alpha_R$ in the compressively strained system by performing band structure calculations along the path perpendicular to the direction of polar displacements $(\bar{X} \rightarrow \Gamma \rightarrow X)$. The intersection point of the bands is set to zero and the curves are fit to parabolas. The data and curve fits of the lower, middle, and upper $t_{2g}$ bands for the undoped, compressively strained ground state polar structure are shown in Fig.~\ref{fig:Fig_6_rashba_splitting}(a)-(c). The Rashba parameters extracted for the lower, middle, and upper conduction bands are 12.6, 25.6, and 34.7~meV/\text{\AA}, respectively. 

\begin{figure}[bp]
\centering
\includegraphics[width=0.5\textwidth]{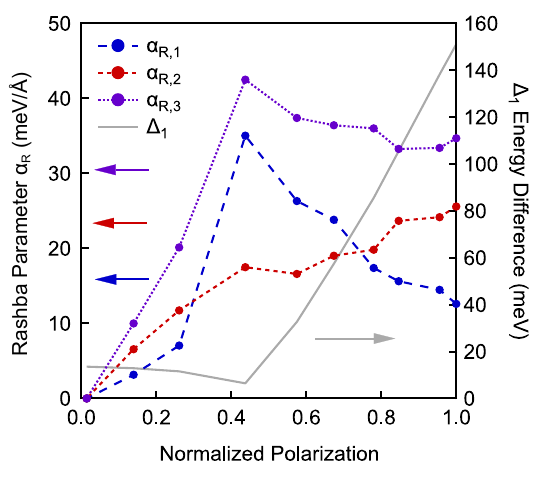}
\caption{\textbf{Rashba parameters.} $\alpha_R$ versus normalized polarization for the lower ($\alpha_{R,1}$), middle ($\alpha_{R,2}$), and upper ($\alpha_{R,3}$) conduction bands. In the limit where the polar distortion is suppressed, the Rashba splitting goes to zero. The energy difference between the lower and middle conduction bands ($\Delta_1$) is plotted on the right axis to demonstrate that the minimum energy difference corresponds to the maximum Rashba parameter.}
\centering
\label{fig:Fig_7_Rashba}
\end{figure} 

Figure~\ref{fig:Fig_7_Rashba} shows the dependence of the Rashba parameters for the lower $(\alpha_{R,1})$, middle $(\alpha_{R,2})$, and upper $(\alpha_{R,3})$ $t_{2g}$ conduction bands, respectively. A larger Rashba parameter corresponds to a larger Rashba splitting within a given band. For the ground state polarization, the Rashba parameter decreases going from the upper to lower energy band. As the polar distortion is suppressed, $\alpha_{R,1}$ and $\alpha_{R,3}$ increase until they reach maxima at $n \approx 0.045$, and then decrease rapidly. The energy gap between the middle and lower energy bands is plotted on the right axis, demonstrating that the maximum values of $\alpha_{R,1}$ and $\alpha_{R,3}$ correspond to the minimum energy difference. The magnitude of $\alpha_{R,2}$, in contrast, decreases monotonically as the polar order is suppressed. This is in general agreement with previous work finding a maximum Rashba parameter close to the paraelectric phase boundary~\cite{varignonUnexpectedCompetitionFerroelectricity2024}. 

A previous study found the Rashba coupling in SrTiO$_3$ to be insufficient to support the experimental superconducting transition temperature, but argued that since $\lambda \propto \omega_{TO}^{-2}$, the superconducting coupling constant diverges as the system approaches the quantum critical point $\omega_{TO} \rightarrow 0$, and that tetragonal domains of AFD order may allow for variations of carrier concentration or the softness of the mode, which may in turn lead to filamentary superconductivity~\cite{gastiasoroTheorySuperconductivityMediated2022}. In the strained system, however, superconductivity is enhanced deep within the polar phase~\cite{russellFerroelectricEnhancementSuperconductivity2019}, where the phonon frequency has hardened to a finite value, as we saw in our Langevin simulations (Fig.~\ref{fig:Fig_4_Langevin}). Furthermore, both experimental and computational work has shown the AFD order to be single-domain in the compressively strained system~\cite{zhuCoexistenceAntiferrodistortivePolar2024}, meaning that the filamentary superconductivity would not likely occur, and could not enhance or account for a strong Rashba coupling. 

\begin{figure*}[htbp]
\includegraphics[width=1\textwidth]{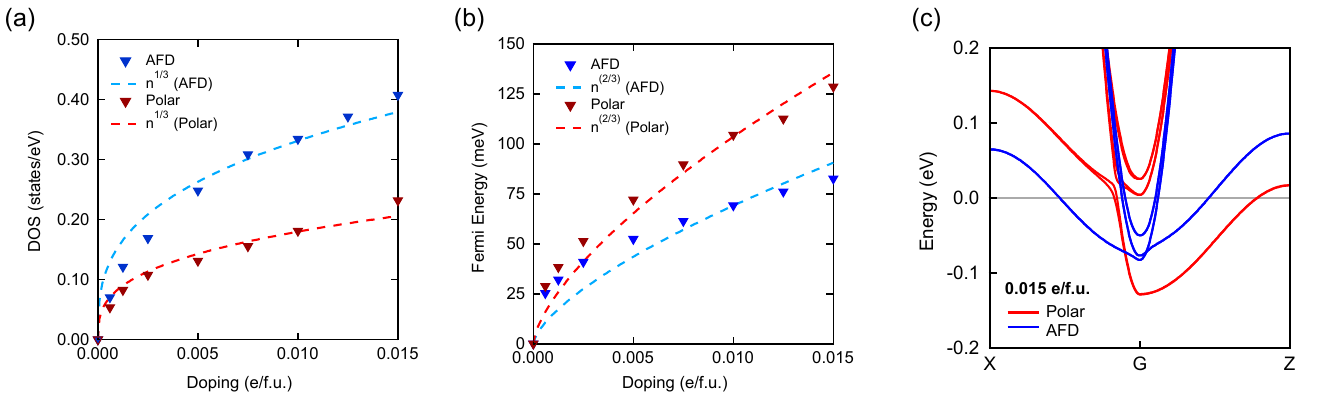}
\centering
\caption{\textbf{Density of states at the Fermi level.} (a)~Density of states versus doping for the strained polar (red) and unstrained AFD (blue) structures. The dashed lines are fits to an idealized $n^{1/3}$ doping dependence. (b)~Fermi energy versus doping for the strained polar (red) and unstrained AFD (blue) structures. The dashed lines are fits to an idealized $n^{2/3}$ doping dependence. (c)~Band structures for the strained polar (red) and unstrained AFD (blue) structures at a doping level of 0.015~e/f.u., corresponding to the highest carrier concentration in the other panels. At this doping, the polar structure is still within the single-band limit, while the AFD structure is in the three-band regime.}
\label{fig:Fig_8_DOS}
\end{figure*} 

\subsection{Density of States}

In conventional BCS theory, the density of states at the Fermi level has a strong influence on the superconducting critical temperature, together with the electron-phonon coupling constant. The total density of states for the unstrained AFD I4/mcm structure is compared to that of the compressively strained polar I4cm structure in Fig.~\ref{fig:Fig_8_DOS}(a). The dashed lines are fits of the data to an $n^{1/3}$ doping dependence expected from a simplified band ellipsoid model, as detailed in the Supplemental Materials~\cite{SeeSupplementalMaterial}. The Fermi energies shown in Fig.~\ref{fig:Fig_8_DOS}(b) are defined relative to the bottom of the lowest conduction band. The band dispersions for the I4cm (red) and I4/mcm (blue) structures plotted in Fig.~\ref{fig:Fig_8_DOS}(c) show that at the highest doping level, the polar structure is still within the single-band limit, while the unstrained AFD structure is in the three-band regime. Here, the density of states is approximately twice as high for the AFD structure as for the polar structure. The single-band density of states is likely lower for the tetragonal structure, as shown in the Supplemental Materials~\cite{SeeSupplementalMaterial}. A higher density of states at the Fermi level within the single, lowest energy band for the polar structure could be one reason for enhanced superconductivity in the polar phase, if pairing occurs only within the lowest-energy conduction band. 

\begin{figure}
\centering
\includegraphics[width=0.4\textwidth]{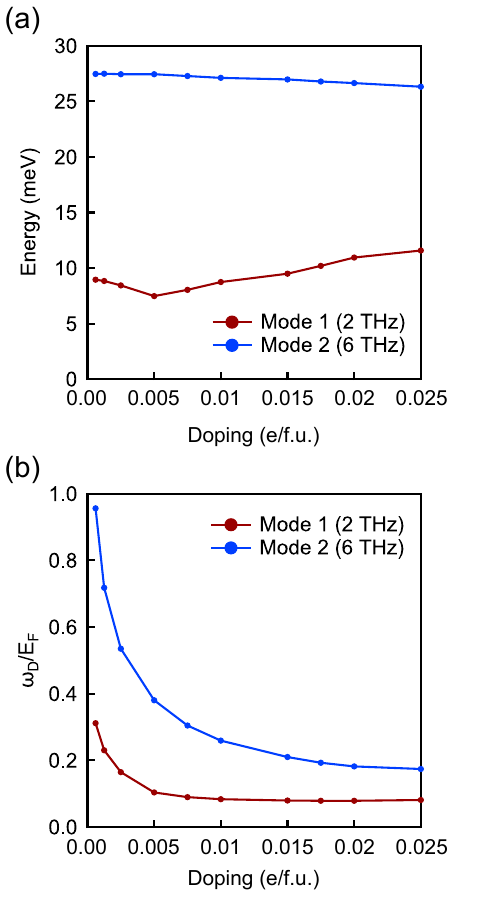}
\caption{\textbf{Phonon energies and Migdal ratios.} (a)~Energy of the two lowest polar phonon modes versus doping. (b)~Migdal ratio of the two lowest polar phonon modes versus doping.}
\label{fig:phonon_energies}
\end{figure}

\subsection{Phonon Energy}

In the conventional phonon-mediated Cooper pairing described by BCS theory, the Midgal ratio $(\omega_D / E_F)$ must be much less than one to justify ignoring the Coulomb repulsion and achieve a net attractive interaction between paired electrons. The doping dependence of the energies of the two lowest polar phonon modes and their corresponding Migdal ratios are shown in Fig.~\ref{fig:phonon_energies}. Phonon frequencies were calculated using density functional perturbation theory, and the Fermi energies are the same as those extracted from the density of states calculations described in the previous section. Both polar modes remain within the adiabatic regime $(\omega_D/E_F) < 1$. The Midgal ratio decreases with doping because the increase in the Fermi energy far outweighs any changes in the phonon frequency. SrTiO$_3$ remaining within the adiabatic regime could suggest that the TO mode plays a role in Cooper pairing. However, the specific mechanism by which the electron density couples to this mode has not yet been elucidated, since Rashba coupling and other phonon mechanisms do not appear to be strong enough to mediate pairing within the polar phase~\cite{gastiasoroTheorySuperconductivityMediated2022}.

\section{Conclusion and Outlook}

In conclusion, we have successfully incorporated the effects of doping into a minimal free energy model of polar order in compressively strained SrTiO$_3$. We find that, in agreement with experimental results, electron doping suppresses the polar transition temperature and the magnitude of the polar order parameter, and makes the formation of polar nanodomains unfavorable. Through Langevin dynamical simulations, we incorporated the effects of thermal disorder to calculate the temperature-dependent phonon spectral function, demonstrating the thermal broadening of spectral lines, a softening of the lowest polar phonon mode at the polar transition temperature, and a subsequent hardening of the mode to a finite frequency within the ordered state, in agreement with complementary Monte Carlo simulations.

In addition, we have calculated the electronic band structure for the strained polar and unstrained AFD tetragonal phases over a range of doping levels. We find that the polar distortion significantly increases the energy splitting between the lowest energy band and the upper bands. Our density of states calculations indicate that, in the limit of pairing within a single-band, an increased density of states could result in enhanced superconductivity in the polar phase. Experimental investigations of the Fermi energy, density of states, and the precise location of the Lifshitz transitions in the strained system could be useful in corroborating these results.

We have also investigated the doping dependence of the Rashba parameter in the polar phase and find that the Rashba splitting is smallest for the lowest energy band, which is likely the most relevant to superconductivity in the polar structure, where the single-band regime is present up to high doping levels. Our conclusion is that Rashba coupling alone may not be sufficiently strong to justify superconductivity in the polar phase. 

We calculated the Migdal ratio over a range of doping levels for the two lowest energy polar modes, whose DFT frequencies agree with the Langevin simulations of the phonon spectral function.  We find the Migdal ratio to be within the adiabatic limit throughout the relevant doping range. It should be noted, however, that the eigenvectors and frequencies of the polar modes may differ from experiment. Also, the level of doping required to completely suppress the polar distortion in DFT is much greater than that of experiments, which could influence the doping dependence of the phonon frequencies.

The techniques implemented here demonstrate the augmentation of first-principles calculations to incorporate the effects of temperature and disorder, and may be applied to other systems with similar success. For example, many models that use Landau theory and DFT calculations do not always take into account random spatial fluctuations of order parameters, and are therefore unable to simulate domain structure. Furthermore, using a simplified free energy model to calculate zero-temperature phonon dispersions can be implemented in material systems where calculating the full dispersion may be too computationally expensive. We have also demonstrated that thermal effects and structural disorder can be incorporated into phonon spectral function models using Langevin dynamics. Our methods can very easily be mapped to other perovksite oxides with similar structural degrees of freedom.

The majority of proposed theoretical frameworks for superconductivity in SrTiO$_{3}$ require proximity to the quantum critical point to justify pairing. We emphasize that in the compressively strained system, enhanced superconductivity is observed far from the quantum critical point, and any complete description of superconductivity in SrTiO$_{3}$ should account for pairing deep within the polar phase in the absence of fluctuations. While most computational studies focus on the bulk or tensile strained systems in which quantum critical fluctuations may be present, our work provides a comprehensive study of the compressively strained system which is far from the quantum critical point. Based on our calculations, we encourage experimental measurements of the phonon frequency and Fermi energy in compressively strained SrTiO$_{3}$, as well as the development of theories of microscopic pairing mechanisms which are not dependent on the presence of quantum fluctuations.

\section*{Acknowledgments}

We would like to thank Susanne Stemmer and Yubi Chen for helpful discussions. This work was supported by the National Science Foundation (NSF) under Grant No.~DMR-2140786. Use was made of computational facilities purchased with funds from the NSF (CNS-1725797) and administered by the Center for Scientific Computing (CSC). The CSC is supported by the California NanoSystems Institute and the Materials Research Science and Engineering Center (MRSEC; NSF DMR-2308708) at UC Santa Barbara. 


%

\end{document}